\documentclass[a4paper,11pt,usenatbib]{article}
\pdfoutput=1 

\usepackage{jcappub} 

\usepackage[T1]{fontenc} 
\usepackage{lineno}
\usepackage{natbib}
\usepackage{hyperref}
\usepackage{verbatim}

\title{\boldmath Searching for Gamma-Ray counterparts to Gravitational Waves from merging binary neutron stars  with the Cherenkov Telescope Array}

\author[a,b]{B. Patricelli,}
\author[a,c]{A. Stamerra,}
\author[b,d]{M. Razzano,}
\author[e]{E. Pian,}
\author[b]{G. Cella}

\affiliation[a]{Scuola Normale Superiore, \\Piazza dei Cavalieri, 7, 56126 Pisa, Italy}
\affiliation[b]{INFN - Sezione di Pisa, \\Largo B. Pontecorvo, 3, 56127 Pisa, Italy}
\affiliation[c]{INAF - Osservatorio Astronomico di Monte Porzio, \\Monte Porzio (RM), Italy}
\affiliation[d]{Dipartimento di Fisica, Universit\`a di Pisa, \\Largo B. Pontecorvo, 3, 56127 Pisa, Italy}
\affiliation[e]{INAF - Institute of Space Astrophysics and Cosmic Physics, \\Via Gobetti 101, I-40129 Bologna,Italy}

\emailAdd{barbara.patricelli@pi.infn.it}
\emailAdd{antonio.stamerra@sns.it}

\abstract{
The merger of binary neutron star (BNS) systems are predicted to be progenitors of short gamma-ray bursts (GRBs); the definitive probe of this association came with the recent detection of gravitational waves (GWs) from a BNS merger by Advanced LIGO and Advanced Virgo (GW170817), in coincidence with the short GRB 170817A observed by Fermi-GBM and INTEGRAL. Short GRBs are also expected to emit very-high energy (VHE, > 100 GeV) photons and VHE electromagnetic (EM) upper limits have been set  with observations performed by ground-based gamma-ray detectors and during the intense EM follow-up campaign associated with GW170817/GRB 170817A.
In the next years, the searches for VHE EM counterparts will become more effective thanks to the Cherenkov Telescope Array (CTA): this instrument will be fundamental for the EM follow-up of transient GW events at VHE, owing to its unprecedented sensitivity, rapid response (few tens of seconds) and capability to monitor large sky areas via survey-mode operation. 

We present a comprehensive study on the prospects for joint GW and VHE EM observations of merging BNSs with Advanced LIGO, Advanced Virgo and CTA, based on detailed simulations of the multi-messenger emission and detection. We propose a new observational strategy optimized on the prior assumptions about the EM emission. The method can be further generalized to include other electromagnetic emission models. According to this study CTA will cover most of the region of the GW skymap for the intermediate and most energetic on-axis GRBs associated to the GW event.
We estimate the expected joint GW and VHE EM detection rates and we found this rate goes from 0.08  up to 0.5 events per year for the most energetic EM sources.
}


\begin{document}
\maketitle
\flushbottom


\section{Introduction}\label{sec:intro}

On 2017, August 17 a binary neutron star (BNS) coalescence was observed through gravitational waves (GWs) by Advanced LIGO \cite{2010CQGra..27h4006H,2015CQGra..32g4001T} and Advanced Virgo \cite{2015CQGra..32b4001A}: GW 170817 \cite{2017PhRvL.119p1101A}; approximately $\sim$ 1.7 s after the merger time, a weak short gamma-ray burst (GRB) has been observed by \emph{Fermi}-GBM \cite{2017ApJ...848L..14G} and INTEGRAL \cite{2017ApJ...848L..15S}. This joint observation provided the first direct evidence that at least a fraction of BNSs are progenitors of short GRBs. The intense electromagnetic (EM) follow-up campaign performed after the coincident GW and gamma-ray detection also allowed the identification of an optical/infrared counterpart to the GW event (the so called ``kilonova''), located at 10'' from the center of the galaxy NGC 4993 (see e.g. \cite{2017Natur.551...67P}, \cite{2017ApJ...848L..12A} and references therein). Furthermore, also an X-ray and a radio counterpart have been detected  \cite{2017Natur.551...71T,2017Sci...358.1579H}; the detection of these X-ray and radio emissions is still ongoing and different models have been proposed to interpret the data, such as emission from an off-axis, structured jet or a cocoon shock breakout (see, e.g., \cite{2018Natur.554..207M,2018ApJ...856L..18M,2018arXiv180106164D}).


High-energy (HE, > 100 MeV) and very-high-energy (VHE, > 100 GeV) gamma-ray detectors (\emph{Fermi}-LAT, HAWC, H.E.S.S.) also followed-up GW 170817, but no EM counterpart has been found \cite{2017ApJ...848L..12A}. Specifically, \emph{Fermi}-LAT was not collecting data at the time of the GW trigger due to a passage through the South Atlantic Anomaly, therefore it was not possible to place constraints on the existence of HE emission associated with the moment of the BNS coalescence. A search has been performed on longer time scales, but no candidate EM counterpart was detected on timescales of minutes, hours, or days after the GW detection \cite{2017arXiv171005450F}. H.E.S.S. observations started only $\sim$ 5 hours after GW170817. The  monitoring campaign was extended over several days and no significant gamma-ray emission has been found. However, the derived upper limits on the VHE gamma-ray flux allowed to constrain non-thermal, HE emission following the merger of a BNS \cite{2017ApJ...850L..22A}.

Searches for high-energy neutrinos from the BNS merger in the GeV - EeV energy range have also been performed using the ANTARES, IceCube, and Pierre Auger Observatories. No neutrinos directionally coincident with the source were detected within $\pm$ 500 s around the merger time, as well as within a 14 day period after the merger; additionally, no MeV neutrino burst signal was detected \cite{2017ApJ...850L..35A}.

In the next years, Advanced LIGO and Advanced Virgo will be upgraded, and will progressively increase their sensitivity up to a factor of ten with respect to the initial LIGO \cite{2009RPPh...72g6901A} and Virgo \cite{2012JInst...7.3012A}; specifically, the final design configuration is expected to be achieved in 2020 and 2021 by Advanced LIGO and Advanced Virgo respectively \cite{2017arXiv1304.0670L}. Many other BNS GW detections are expected in the near future (0.1-200 events/year \cite{2017arXiv1304.0670L}), and multi-messenger astronomy will be key to further probe the rich physics of these transient phenomena.  

One of the next challenges for multi-messenger astronomy will be the detection of VHE gamma rays associated with GW signals. In fact, short GRBs are expected to emit also VHE photons; however, until now the detection of VHE photons remained elusive, possibly due to the limited sensitivity and high-energy threshold of current Cherenkov detectors (see, e.g., \cite{2005ApJ...633.1018P}). Furthermore, VHE photons are attenuated due to interactions with the extragalactic background light (EBL); therefore, their detection is possible only for very bright and nearby bursts, such as the ones possibly associated with sources detectable by Advanced LIGO and Advanced Virgo. 

In $\sim$ 2020 the Cherenkov Telescope Array (CTA, see \cite{2013APh....43....3A}), an advanced, next generation ground-based facility will become operative. Thanks to its unprecedented sensitivity, its rapid slewing capabilities and its large field-of-view (FOV), CTA  represents an ideal instrument to detect the VHE emission from short GRBs associated to GW events. Furthermore, CTA will have a coincident observational schedule with GW detectors at design sensitivity and third generation of ground-based interferometer \cite{2010CQGra..27h4007P,2011CQGra..28i4013H,2015PhRvD..91h2001D}, when several GW triggers are expected to be shared with the astronomical community. 

Despite the large FOV of CTA, the poor sky localization of the GW events (from tens to hundreds of square degrees, see, e.g., \cite{2016JCAP...11..056P,2017arXiv1304.0670L}) requires the definition of an observational strategy for the search of the EM counterpart, through an optimized scan of the sky map provided by the GW events. Recently, \cite{2014MNRAS.443..738B} explored the feasibility of following up GW events over these large sky areas and possibly detect the short GRBs associated with the GW events with CTA. They considered CTA operating in survey mode and assumed the same observing time for each consecutive CTA observation; this observing time has been estimated considering two possible values for the GW sky localization area (200 deg$^2$ and 1000 deg$^2$) and taking into account the total duration of the GRB emission (1000 s). They found that short GRBs with emission extending up to 100 GeV can be detected by CTA if observations are delayed no more than 100 s with respect to the GRB onset, while short GRBs with emission at lower energies can only be detected with lower delay times ($<$ 10 s).

In this work we investigate the prospects for joint GW and VHE EM observations with Advanced Virgo, Advanced LIGO, and CTA based on detailed simulations of BNS mergers accompanied by short GRBs. Our study explores different ranges of source parameters that, either known or assumed,  allow us to optimize the observational strategy and to increase the probability of the detection of VHE EM counterparts to transient GWs with CTA. The main novelty of our approach is that the observing time is not set to a single value for all the observed fields, but it is estimated for each source (and for each consecutive observed field) taking into account the fading of the EM emission, as well as the CTA sensitivity for different integration times. 

The work is organized as follows: in Sec. \ref{sec:gevgrbs} we review the currents status of GRB observations at GeV energies and in Sec. \ref{sec:cta} we summarize the CTA characteristics. In Secs. \ref{sec:BNSmerger} and \ref{emmodeling} we describe the sample of simulated BNS systems that we use in this work, and how we simulated the associated EM signal. In Secs. \ref{sec:strategy} and  \ref{ctadetection} we present the observational strategy proposed to optimize the probability of detection of the EM counterparts. In Sec. \ref{sec:test} we show the results obtained applying the strategy to a specific test case. In Sec. \ref{sec:results} we discuss our overall results and in Sec. \ref{sec:concl} we present our conclusions.

\section{HE and VHE emission from short GRBs}\label{sec:gevgrbs}

GRBs are predicted to emit VHE gamma rays in the framework of the fireball model through different possible mechanisms such as the leptonic Synchrotron Self-Compton (SSC, see, e.g., \cite{1999ApJ...512..699C,2001ApJ...559..110Z,2014ApJ...787..168V}), as well as hadronic processes involving proton-synchrotron radiation (see, e.g., \cite{2010OAJ.....3..150R,2010ApJ...724L.109R}) and photo-hadronic interactions resulting in cascade gamma-ray production \cite{1998ApJ...499L.131B}. Detecting VHE photons from GRBs could help us to better understand the physical composition of the outflow, the radiation mechanisms, and the underlying physical processes.

From the observational point of view, prior to the launch of the \emph{Fermi} mission there was just a limited knowledge about GRB emission at HE. For instance, EGRET detected a 18 GeV photon from the long GRB 940217 \cite{1994Natur.372..652H} and HE emission (up to 200 MeV) from the long GRB 941017 \cite{2003Natur.424..749G}. With \emph{Fermi}, the number of GRB detections at high energies is constantly increasing. Up to now, dozens of GRBs with high energy emission have been detected, including GRB 130427A, with the highest energy photon at 95 GeV (128 GeV in the rest frame, see \cite{2014Sci...343...42A}); among these souces, six are short GRBs, two with emission above 1 GeV: GRB 081024b (highest energy $\sim$ 3 GeV \cite{2010ApJ...712..558A})  and GRB 090510 (highest energy $\sim$ 30 GeV \cite{2010ApJ...716.1178A}). At very high energies, only a hint of TeV emission was detected by Milagrito (500 GeV-20 TeV) from the long GRB 970417A \cite{2000ApJ...533L.119A}. Searches for TeV emission from GRBs with the current Imaging Atmospheric Cherenkov Telescopes (IACTs) have only allowed upper limits on the VHE emission from GRBs (see, e.g., \cite{2007ApJ...667..358A,2011ApJ...743...62A}) to be set; only upper limits have been obtained also with water Cherenkov detectors such as Milagro \cite{2012ApJ...753L..31A} and HAWC \cite{2015ApJ...800...78A}.

\section{The Cherenkov Telescope Array}\label{sec:cta}
CTA\footnote{\url{https://www.cta-observatory.org/}} is an international project aiming to build and operate a new observatory for VHE gamma rays \cite{2013APh....43....3A}. In its current design, CTA will be composed of two arrays, one in the northern hemisphere and one in the southern hemisphere, which together will provide full-sky coverage. CTA will be an order of magnitude more sensitive and will have a greater energy coverage (from a few tens of GeV to above 100 TeV) with respect to current IACTs; it will also have a better angular and energy resolutions. The two arrays will consist of a combination of large (LST), medium (MST) and small (SST) size telescopes, covering different  energy ranges: 20 GeV - 100 GeV, 100 GeV - 10 TeV and 10 TeV - > 100 TeV respectively. It is expected to be completed by 2024, but operations are expected to start by 2020, so it will have a coincident observational schedule with GW detectors at design sensitivity \cite{2017arXiv1304.0670L}.  The follow-up of GW transients will start in the early phase of CTA construction according to the core scientific program proposed by the CTA consortium \cite{2017arXiv170907997C}.

\section{BNS mergers and their GW detections}\label{sec:BNSmerger}
The sample of BNS mergers and their GW detection used in this work is based on the simulations produced by \cite{2016JCAP...11..056P}, who designed a specific Montecarlo simulation pipeline for the BNS multimessenger emission and detection by GW interferometers. In particular, they created a realistic ensemble of BNS merging systems in the Local Universe up to 500 Mpc, that is consistent with the expected horizon\footnote{The horizon is the maximum distance at which the interferometers can detect an optimally located, optimally oriented BNS merger.} for BNS mergers of Advanced Virgo and Advanced LIGO in their final configuration \cite{2016LRR....19....1A}. They also 
simulated the associated GW emission and detection by the 2nd generation interferometers in different configurations, taking into account their duty cycle. Specifically, for each merging BNS system, they simulated the expected GW inspiral signals using the ``TaylorT4'' waveforms (see, e.g., \cite{2009PhRvD..80h4043B}). After the GW signals have been simulated, they convolved them with the GW detector responses, using the sensitivity curves of Advanced LIGO and Advanced Virgo reported in \cite{2017arXiv1304.0670L} for the 2016-2017 and the final design configuration. The data obtained in this way were then analyzed with the matched filtering technique \cite{wainstein63,2014PhRvD..90h2004D,2015PhRvD..91d2003V,2016CQGra..33q5012A,2016CQGra..33u5004U,2012ApJ...748..136C,2017PhRvD..95d2001M,2017ApJ...849..118N}. Finally, for each GW simulated candidate they estimated the associated sky localization with BAYESTAR, that is a rapid Bayesian position reconstruction code that computes source location using the output from the detection pipelines \citep{2014ApJ...795..105S}. 

We used this ensemble of BNS merging systems and their simulated GW emission and detection, with focus on the final design configuration of Advanced LIGO and Advanced Virgo and considering an 80 \% independent duty cycle for each interferometer. Specifically, we used the sample produced considering, among the various population synthesis model by \cite{2012ApJ...759...52D}, those predicting the highest BNS merger rate: 830 Gpc$^{-3}$ yr$^{-1}$ \cite{2016JCAP...11..056P}; however, much higher values are also allowed, according to the current merger rate estimated after the BNS detection by Advanced LIGO and Advanced Virgo (320 - 4740 Gpc$^{-3}$ yr$^{-1}$, see \cite{2017PhRvL.119p1101A}).


\section{Simulation of short GRBs}\label{emmodeling}
As done in \cite{2016JCAP...11..056P}, we assumed that all the BNS mergers are associated with a short GRB; we also assumed that the GRB jet is beamed perpendicular to the plane of the binary's orbit (i.e., that the angle of the observer with respect to the jet is equal to the inclination angle of the BNS system $\theta$, see e.g. \cite{2013MNRAS.430.2121P}). Due to the latency needed to send the GW alerts, in this work we only focus on the afterglow emission, that can potentially be detected also at later times; we don't consider here the possible serendipitous detection of a GRB prompt emission.

\subsection{Modeling the GeV emission of short GRBs}\label{sec:emmodel}
We assumed that all short GRBs have an afterglow emission at high energies and we investigated the possibility to detect it with CTA. We assumed a fiducial value of the jet opening angle $\theta_j$=10$^\circ$ (see \cite{2014ApJ...780..118F,2015ApJ...813...64D}) and we only focused on the on-axis GRBs, i.e., on GRBs for which the angle between the line of sight and the jet axis is $\theta < \theta_j$. The number of GW detected events fulfilling this requirement is 960, spanned over 1000 simulations of 1 year each \cite{2016JCAP...11..056P}.

We simulated the GeV afterglow emission using GRB 090510 as a template, since this is the only short GRB to show an extended emission ($\sim$ 200 s) up to GeV energies (up to $\sim$ 30 GeV) \citep{2010ApJ...716.1178A}. Since to date there has been no confirmed TeV photon detection from short GRBs, we simply extrapolated the \emph{Fermi}-LAT observations to higher energies. 
In the synchrotron interpretation of the HE emission a cut-off could be present at high energies (see, e.g., \cite{1996ApJ...457..253D,2014MNRAS.443..738B}); we therefore considered a power-law with exponential cut-off spectrum. 
Specifically, we assumed a spectral index \mbox{$\beta$=-2.1} (see \cite{2010ApJ...709L.146D,2010A&A...510L...7G}) and two different values for the cut-off energy: 30 GeV and 100 GeV. Then, we assumed the same temporal behaviour of the emitted flux observed for GRB 090510, i.e. a smoothly broken power law: \mbox{$F(\rm{t})=A \frac{(\rm{t/t_{peak}})^{\alpha}}{1+(\rm{t/t_{peak}})^{\alpha+\delta}}$} (see \cite{2016JCAP...11..056P} for details). The observed  VHE light curve has been corrected for the distance of the sources with respect to GRB 090510, whose redshift is z=0.903$\pm$0.001 \citep{2010A&A...516A..71M}. A further correction has been applied to take into account that GRB 090510 is a uniquely bright burst: with a prompt emission isotropic energy $E_\gamma=3. 5\times 10^{52}$ erg (excluding the LAT component, see e.g. \cite{2010A&A...510L...7G}), it is in fact among the most energetic GRBs ever observed. To do this correction, we simply re-scaled the GeV light curve for $E_\gamma$ (see \cite{2016JCAP...11..056P} for details). 

The observed duration of the VHE extended emission of GRB 090510 has been of $\sim$ 200 s; however, this could be due to the limited sensitivity of \emph{Fermi}-LAT; if GRB 090510 occurred in the local universe, its flux could have been intense enough to be detectable for more than the observed duration. Therefore, in this work we put no limits on the duration of the VHE extended emission of the simulated GRBs. However, we considered a maximum total observing time for the EM follow-up with CTA of 10$^4$ s, i.e., approximately 3 hours.

\subsection{The EBL absorption}
The Extragalactic Background Light (EBL) is a diffuse and nearly isotropic background of infrared, optical and ultraviolet radiation originating from both resolved and unresolved extragalactic sources. The interaction of VHE photons with the EBL produces e$^+$-e$^-$ pairs, with the consequent attenuation of the $\gamma$-ray flux (see \cite{2013APh....43..112D} for a review). 

The EBL absorption increases with the redshift of the source and  the $\gamma$-ray energy. according to the current EBL models (see, for example, \cite{2011MNRAS.410.2556D}), photons up to energies of $\sim$ 100 GeV emitted by a source located at a distance of $\sim$500 Mpc are not significantly affected by pair annihilation with the EBL\footnote{The maximum decrease in the flux is of the order of a few percent.}, therefore we neglected the EBL absorption when simulating short GRB with an exponential cut-off spectrum (see Sec. \ref{sec:emmodel}). 

\section{An optimized observation sequence for the follow-up of GW alerts}\label{sec:strategy}

The uncertainty in the sky location of the GW event requires the definition of an observational strategy for the search of the EM counterpart with pointed instruments through a scan of the sky map provided by the GW event. Among possible strategies, there is the selection of the fields to be observed based on the surface density of nearby galaxies (see, e.g., \cite{2016ApJ...820..136G}). For instance, in the EM follow-up of GW170817 several teams observed previously cataloged galaxies (for example from \cite{2011CQGra..28h5016W}) in the three-dimensional LIGO-Virgo localization taking into account the galaxy stellar mass and star formation rate \cite{2017Sci...358.1556C}. While this strategy is optimal for the closest GW events such as GW170817, this could be not the case for the EM follow-up of more distant  GW sources, due to the incompleteness of the current existing galaxy catalogs. For instance, the GLADE catalog is complete only up to $\sim$ 70 Mpc \cite{2016yCat.7275....0D} and the Gravitational Wave Galaxy Catalog do not extend beyond 100 Mpc \cite{2011CQGra..28h5016W}. When Advanced LIGO and Advanced Virgo reach their final design sensitivity, a much higher completeness will be needed. Therefore, in this work we use a more conservative approach that does not take into account the galaxy distribution. 


For gamma-ray detectors and for IACTs the key factor is the choice of an optimal integration time that combines the possibility of a significant detection in the sky field observed, with a maximal coverage of the GW uncertainty map. Observing the whole skymap with equally long  snapshots maximizes the coverage, but each snapshot could be too short to yield a significant detection, especially considering that the EM emission fades with time.

In case some of the parameters defining the source and its GRB emission (e.g. the distance, or the spectral slope) are known or can be reasonably assumed from modelling, then a different strategy can be used. We propose to optimize the sequence and duration of observations maximising both the probability of a detection in a single snapshot and the coverage of the uncertainty region.  Such strategy can be applied  to whatever telescope with a limited FOV. Similar observing strategies have been implemented in \cite{Salafia:2017rm},  but focussed to the macronova emission expected in the infrared-optical bands.
The observational strategy proposed here is based on the prior knowledge of the transient parameters. Only basic assumptions are done, as described in Sec. \ref{emmodeling}: the source emits a  E$_{\rm iso}$ isotropic energy, with power-law with exponential cut-off spectrum $E^{-\alpha} \exp(-E/E_c)$
; the EM signal in a given energy band fades according to a function $F(t)$ (see Sec. \ref{sec:emmodel}),  starting at the $t_0$ of the burst.   

The expected flux  depends on the intrinsic source emission model, on its distance and on the delay time from the burst before observation is started. The detection is evaluated through the instrument response functions describing the sensitivity of the instrument.
  
The emission is detectable if the integrated flux in the observation time $t^{obs}_j$ of the $j$-th snapshot  is higher than the minimum detectable fluence $F_s(t^{obs}_j)$  (evaluated in the same energy band as $F(t)$) at a given significance level $s$: 
 
 \begin{equation}\label{eq:detCond}
 \int_{T_j}^{T_j+t^{obs}_j} {F(t)dt \ge F_s(t^{obs}_j)};
 \end{equation}
 
 where $T_j$ is the time passed from the burst until the field $j$ is observed. $T_j$  depends on the latency time for receiving the GW alert ($T_{alert}$) and the slewing time of the telescopes ($T_{slew}$): $T_j = T_{alert} + T_{slew} + \sum_{k=1}^{j-1}t_k^{obs} $. Equation \ref{eq:detCond} sets the generic requirement to be fulfilled for the detection.  It contains the source-dependent parameters, and the instrumental and observational parameters. $F(t)$ includes the transient properties, derived from the assumed physical or phenomenological model; hence, this strategy maximises the probability  of detection of EM counterparts with the assumed properties. 

The observational strategy sets the number of snapshots and the duration of each observation, so that the detection for each snapshot $j$ is granted. At the time $T_1$ when first observation can start, an iterative process computes at each step the observation time $t^{obs}_j$ needed to verify the validity of equation \ref{eq:detCond}. For the most general case of a fading lightcurve, a maximum number of snapshots $N_{max}$ exists, after which no detection is possible anymore.
This happens when the flux is too low and observation time too long to grant a detection. The $N_{max}$ regions observed are then superimposed to the GW skymap, to get the coverage. The choice of the direction for each of the $N_{max}$ regions is optimized to have the highest possible integral probability and thus coverage. 
As already mentioned, further optimization criteria can be applied, e.g. superimposing the GW sky localization area to the distribution of galaxies and choosing the regions with highest density of galaxies in the range of distances provided by the GW event, or with a set of preferred host galaxy type; they are not implemented here and will be considered in successive works. The strategy here described will be detailed in Sec. \ref{ctadetection}.

\section{Detecting short GRBs: EM follow-up with CTA}\label{ctadetection}
Observations with CTA are limited by the total amount of observation time dedicated to the survey and by the maximum time after the burst that justifies the search of a fading counterpart. This depends on the model applied for the EM emission, but reasonable assumptions can drive the choice of the observation time for CTA.
In this section the steps of the simulation performed to derive the probability of detection of short GRBs with CTA is outlined.
 



\subsection{The observation latency}\label{sec:latency}
Due to the latency needed to send the GW alerts to astronomers, the starting time of the EM follow-up observations doesn't coincide with the onset of the GRB emission. 

The typical time-scales with which GW alerts have been sent out by the LIGO-Virgo collaboration during O1 and O2 are of the order of tens of minutes (see, e.g., \cite{2017ApJ...848L..12A}), mainly due to the "human event validation" of the GW triggers. In the future, the procedure to sent GW alerts should become automatic, and it will require only a few minutes, i.e.  the time required by online pipelines to report GW candidates (see, e.g., \cite{2016PhRvD..93d2004K,2017PhRvD..95d2001M}).  

In this work we considered a latency to send the GW alert of $T_{\rm alert}$=180 s; to estimate the total latency, we added to $T_{\rm alert}$ the slewing time of the CTA telescopes, estimated to be $T_{\rm slew}$=30 s \cite{2013APh....43..317D}. A further average slewing time of 15 s is considered in case CTA telescopes are moved towards a distant region of the sky with respect to the previous observed ones during the same follow-up campaign, for instance when the GW skymap is composed by two (or more) isolated ``spots'', located in the same hemisphere. We considered as negligible the slewing time of the telescopes between consecutive pointings covering the same ``spot''.
 
\subsection{Tiling the GW sky map}\label{sec:tiling}

We considered an all-sky survey and we constructed a 2D grid of CTA pointings. Specifically, we defined multiple evenly-spaced row of pointings, assuming an angular step between adjacent pointings of 2 deg, that is the maximum step that allows us to provide nearly uniform sensitivity coverage (see \cite{2013APh....43..317D})\footnote{Different pointing strategies can be used, e.g. the divergent mode \cite{2015ICRC...34..725G} or with alternative observations modes \cite{2015APh....67...33S}, but are not considered in this work.}. 
We assumed that for the regions of the sky located at Dec $\geq$ 10 deg observations are performed by the North array, otherwise the CTA South array is considered.

For each simulated event we selected, among all the pointings of the grid, the ones that are within the 90\% credible region (C.R.) of the GW skymap. 



\subsection{The observation times}\label{sec:obstime}
As already explained in Sec. \ref{sec:strategy}, the strategy here proposed for the EM follow-up of GW events is a combination of two factors: the need to observe the field for enough time to be able to detect also the faintest sources, and the need to cover the largest possible area of the GW skymap. 

Since the source is fading, the longer is the delay between the trigger and the starting of the observations, the lower is the flux, so the higher should be the observation time needed to detect it. To estimate the optimal observation time for each of the consecutive observations performed with CTA, we used the CTA sensitivity as a ``guideline'' in the following way. First of all, we estimated the CTA sensitivity as a function of the observing time. To do this, we used \verb|ctools|\footnote{\url{http://cta.irap.omp.eu/ctools/}; in this work we used the version 1.4.0.}: a software package specifically developed for the scientific analysis of CTA data \cite{2016A&A...593A...1K}; the sensitivity has been computed using the \verb|ctools| function \emph{cssens}. We adopted the  provided CTA instrument response functions (IRFs) in {\em fits}  format, equivalent to those for the public array configurations available at the CTA website\footnote{\url{https://www.cta-observatory.org/science/cta-performance/} (version 2015-05-05) corresponding to following layouts: for CTA-North, layout ``2N'', made up of 4 LSTs and 15 MSTs; for CTA-South, layout ``2Q'' composed of 4 LSTs, 24 MSTs and 72 4m Schwarzschild Couder SSTs.}, computed by the CTA consortium from detailed Monte Carlo simulations, in the so-called "Production 2" \cite{2015arXiv150806075H}. The IRFs for the two arrays ``North$\_$0.5'' and ``South$\_$0.5h'' have been generated by assuming 30-min observation of a point source observed at a zenith angle $\theta_{\rm z}$=20$^\circ$ and located at the center of the FOV \cite{2016A&A...593A...1K}. 

We considered the sensitivity to a point source with the same spectral properties of the simulated GRBs (spectral index and cut-off energy) located at the center of the FOV and we choose a detection threshold corresponding to the 5 $\sigma$ post-trials (see Appendix \ref{sec:TS}). We considered the energy range covered by the LSTs and the MSTs: 30 GeV - 10 TeV (nominally, the LSTs are sensitive down to 20 GeV; however, at energies below 30 GeV there is not sufficient information in the current instrument response functions to assure a reliable simulation and analysis in all situations with  \verb|ctools|). Then, for each simulated source and for each CTA consecutive observation (each one starting with an increasing delay with respect to the trigger), we estimate the time we need to observe it in order for its fluence to be equal to the CTA sensitivity for that observing time; to do this, we assume the distance of the source provided by the GW alert. Since the source emission is fading, after some time the source will not be detectable anymore because, even with long observing times, the fluence is below the CTA sensitivity: this will give us an estimate of the maximum number of allowed pointings with CTA, and therefore of the percentage of the GW skymap that can be covered with CTA tiles. When the percentage of the GW skymap that can be covered is lower than 90 \%, the CTA pointings are selected based on the associated GW probability (starting from the ones associated with the highest probability regions).


\section{Testing the observational strategy on a specific case}\label{sec:test}
In the following we show the results obtained by applying the strategy described in the previous sections to a BNS merger from our sample. 
To this scope, we selected from the database produced by \cite{2016JCAP...11..056P} one BNS merger detected by Advanced LIGO and Advanced Virgo and whose inclination angle is less than 10$^\circ$, so that the possible associated GRB is seen on-axis. This source is located at a distance of 215 Mpc; the combined signal-to-noise ratio (SNR) of the GW detection is $\sim$ 18, and the 90 \% C.R. in the sky localization is of $\sim $56 deg$^2$. 

\subsection{Application of the strategy}

\begin{figure}[h!]
\begin{center}
\includegraphics[scale=0.5]{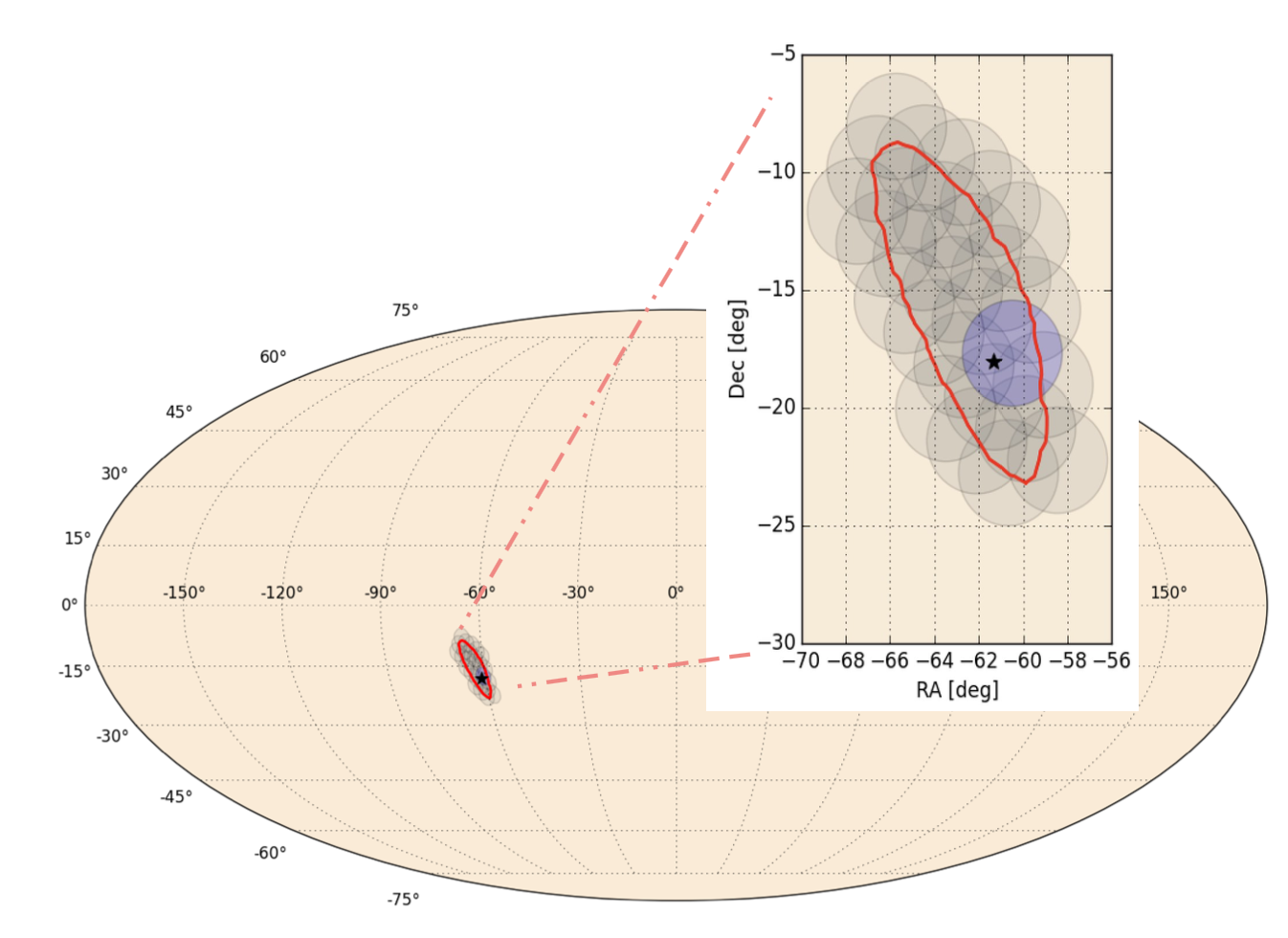}
\caption{Representation of the 90 \% C.R. in the sky localization of the selected GW event (red contour) and the consecutive set of sky areas that can be covered by CTA follow-up observations (gray and blue circles), assuming E$_{\rm iso}$=10$^{51}$ erg and E$_{\rm cut}$=100 GeV. The source position is marked with a black star. The CTA observed field in which the source is detected is represented by a blue circle. For simplicity, the amplitude of each CTA observed area is represented taking into account only the FOV of LSTs. The inset shows a zoom of the GW sky localization area and CTA tilings.}\label{fig:tiling}
\end{center}
\end{figure}

\begin{figure}[h!]
\hspace{-1cm}
\begin{center}
\includegraphics[height=8cm]{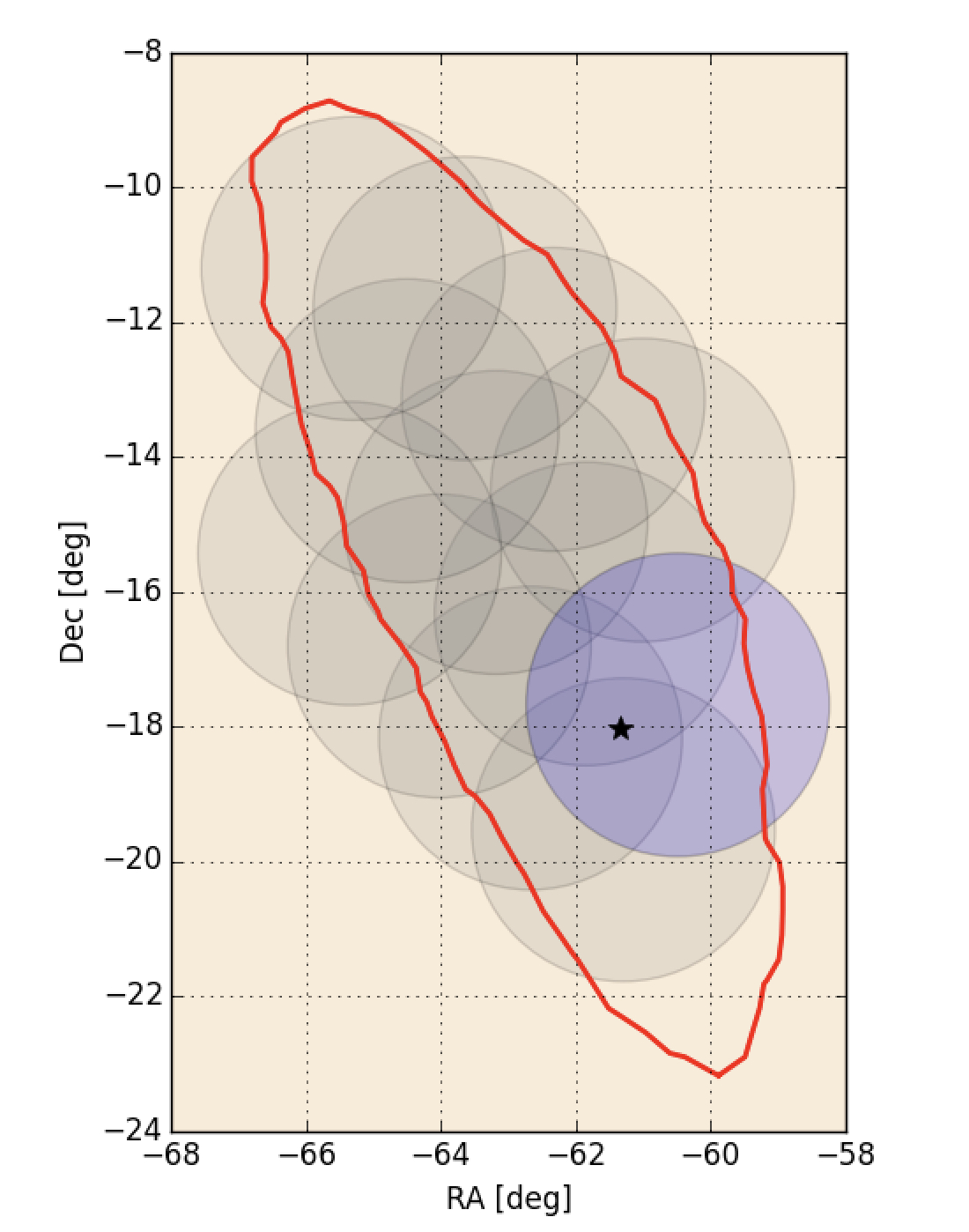}
\caption{Representation of the 90 \% C.R. in the sky localization of the selected GW event (red contour) and the consecutive set of sky areas that can be covered by CTA follow-up observations (circles, as in Fig. \ref{fig:tiling}), assuming E$_{\rm iso}$=10$^{50}$ erg and E$_{\rm cut}$=100 GeV.}\label{fig:tiling2}
\end{center}
\end{figure}

We apply the proposed observational strategy to the selected event. First we assumed an intermediate isotropic energy E$_{\rm iso}$=10$^{51}$ erg and a cut-off energy E$_{\rm cut}$=100 GeV. We found that, with the above assumptions, the EM flux of the source is high enough to allow the coverage of the whole 90 \% credible region (C.R.) of the GW skymap, i.e. the region of the GW skymap enclosing the 90 \% of probability that the source is located there.
The corresponding tiling of the GW skymap is shown in fig. \ref{fig:tiling}. For comparison, we also considered the cases E$_{\rm iso}$=10$^{50}$ erg and a cut-off energy E$_{\rm cut}$=100 GeV; in these cases, only 50 \% C.R. of the GW skymap can be covered (see fig. \ref{fig:tiling2}), but this coverage is still sufficient to detect the source. 

\subsection{Validation of the strategy}
To verify the validity of the proposed strategy to detect short GRBs, we performed accurate simulations of the GRB emission and detection with \verb|ctools|. Specifically, we used the function \emph{ctobssim} to create the event maps for the simulated GRB, assuming a total latency and duration of the observations as estimated with the previously described strategy. We then verify if the source is detected producing the Test Statistic (TS) map (see Appendix \ref{sec:TS}) with the function \emph{cttsmap}: we performed our simulations with a ROI of 2.25$^\circ \times 2.25^\circ$ and a pixel size of 0.2$^\circ$. 

\begin{figure}[h!]
\begin{center}
\includegraphics[height=7cm]{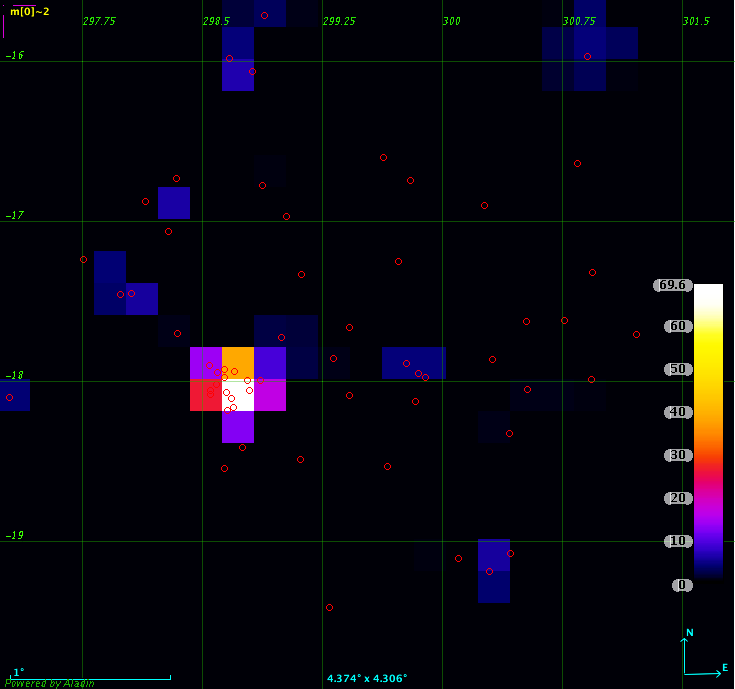}
\caption{TS map for the selected GRB; the superimposed red points represents the event map.}\label{fig:maps} 
\end{center}
\end{figure}

One of the event maps obtained for the simulated GRB as seen by CTA South in the energy range 0.03 - 10 TeV is shown in Fig. \ref{fig:maps}, together with the associated TS map. It can be seen that the TS value is $\sim$ 70: this corresponds to a statistical (post-trial) significance $> 5 \sigma$: therefore, the adopted strategy allows to detect the source. 

\begin{figure}[h!]
\begin{center}
\includegraphics[scale=0.5]{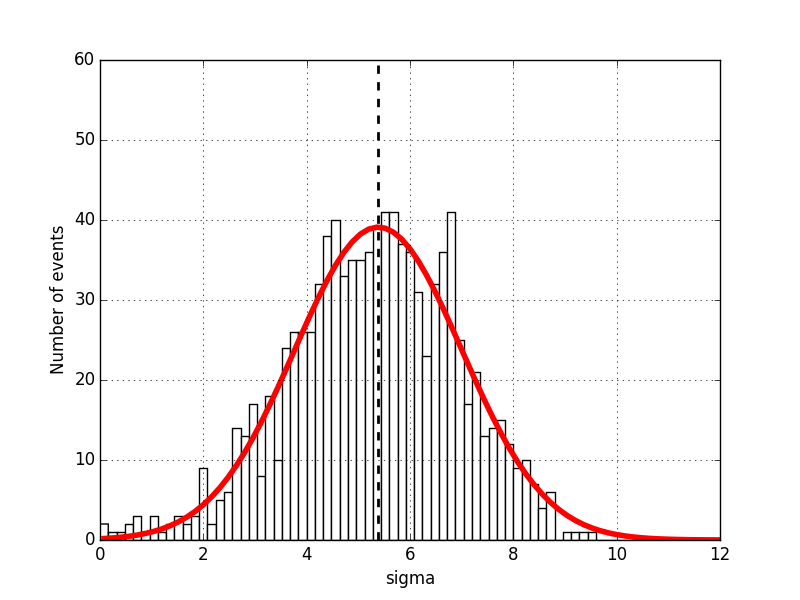}
\caption{Distribution of the statistical significance of the EM detection (black) and best fitting gaussian function (red solid line).}\label{fig:sigma}
\end{center}
\end{figure}

Different observations of a source with a given intrinsic spectrum could yield different statistical significances, due to fluctuations in the number of detected photons. We evaluated these fluctuations repeating the simulation 1000 times with different seeds. For each simulations, we got a statistical significance for the EM detection of the event. The distribution of the statistical significance of the detection for the observed field containing the true source position is shown in Fig. \ref{fig:sigma}. It can be seen that the distribution is well described by a Gaussian with mean (post-trials) $\mu \sim$5 and $\sigma \sim$ 1.3, as expected.

\section{Results and Discussion}\label{sec:results}

\subsection{GW skymaps coverage}

\begin{figure}[h!]
\begin{center}
\includegraphics[height=6cm]{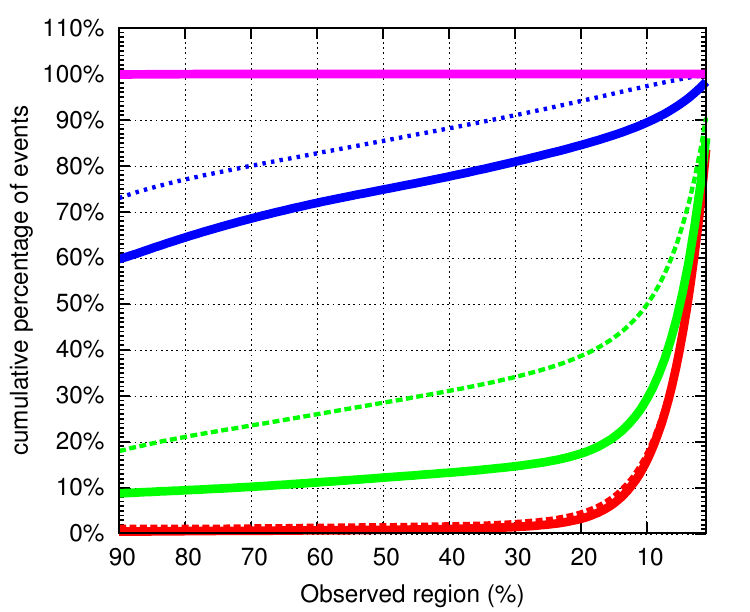}\includegraphics[height=6.5cm]{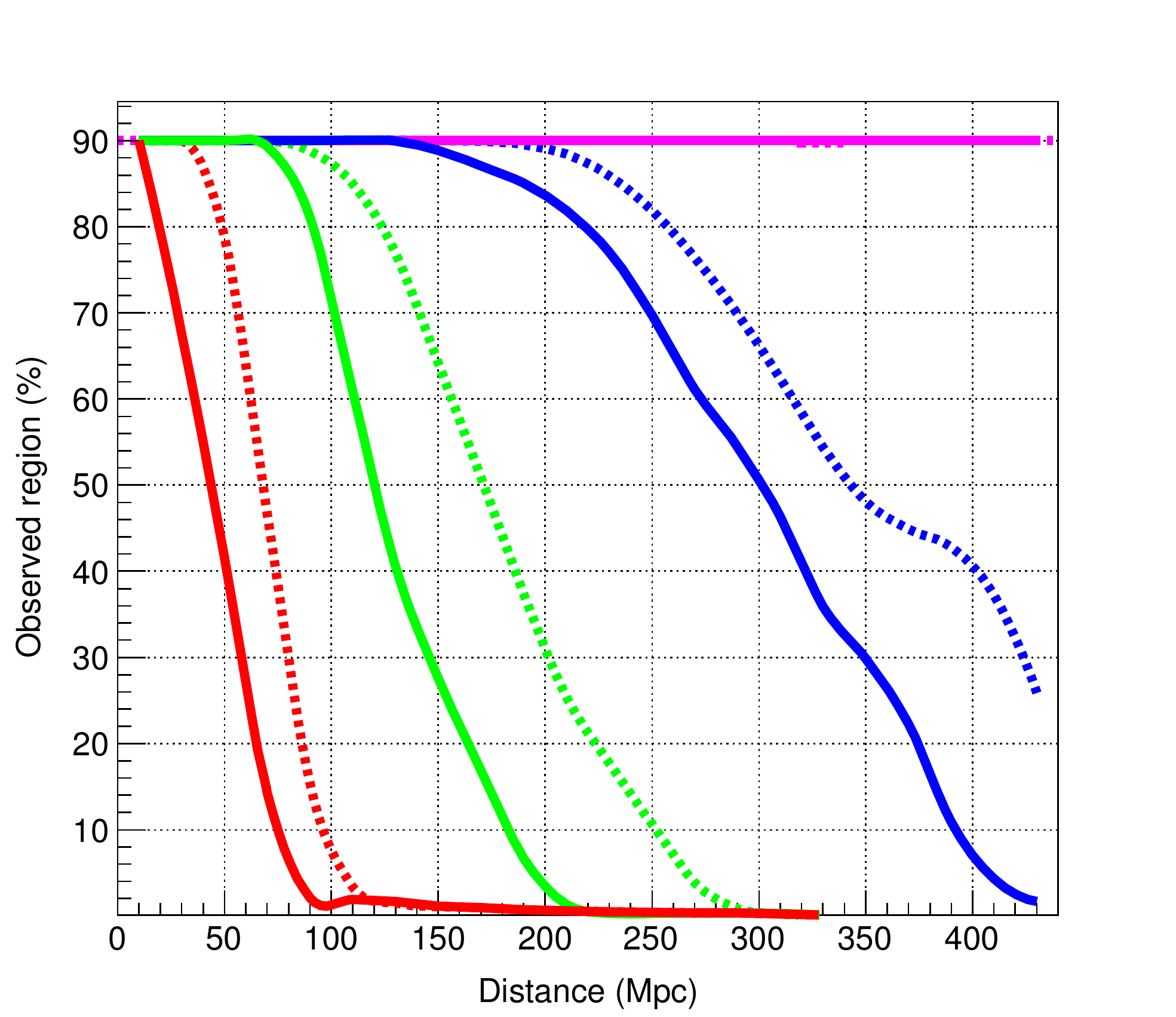}
\caption{Left panel: cumulative histogram of  GW sky localization areas that can be covered with CTA consecutive observations, assuming a cut-off energy of 100 GeV (dotted lines) and 30 GeV (solid lines). The different colours refer to different assumed values for the isotropic energy of the simulated short GRB: 3.5 10$^{52}$ erg (magenta), 10$^{51}$ erg (blue), 10$^{50}$ erg (green) and 10$^{49}$ erg (red). Right panel: GW sky localization areas that can be covered with CTA consecutive observations, as a function of the distance of the BNS systems. The color and line style schemes are the same as in the left panel.}\label{fig:cumarea}
\end{center}
\end{figure}

For each of the simulated events, we used the proposed strategy to determine the percentage of the GW skymap that can be covered with consecutive CTA observations, in order to be able to detect the EM emission. This percentage mainly depends on: a) the source position (distance and sky coordinates), which determine the amplitude of the GW error box and b) the parameters characterizing the EM emission, together with the CTA sensitivity. The cumulative histogram of the GW sky localization areas that can be covered with CTA, taking into account the different assumptions, is shown in Fig. \ref{fig:cumarea} and in Tab. \ref{tab:coverage}.

\begin{figure}[h!]
\begin{center}
\includegraphics[height=6cm]{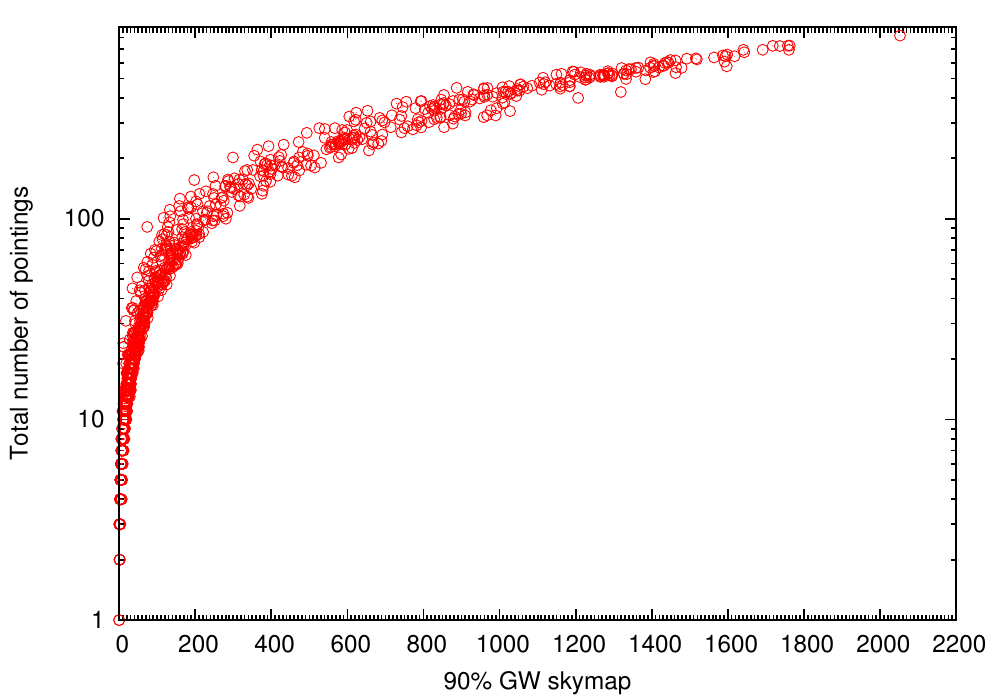}
\caption{Total number of CTA pointings (including both the North and the South array pointings) needed to cover the 90 \% credible region of the GW skymaps, as a fuction of the area corresponding to the 90\% credible region for all the simulated BNS events.}\label{fig:totpointings}
\end{center}
\end{figure}

It can be seen that, when short GRBs as energetic as GRB 090510 are considered, CTA is sensitive enough to allow us to cover the 90 \% C.R. of the GW skymap for $\sim$ 100\% of the simulated events, independently on the cut-off energy and on how large this sky area is (the number of pointings required to cover the 90 \% C.R. of the GW skymap as a function of the size of the corresponding sky area is shown in Fig. \ref{fig:totpointings}).

For an intermediate isotropic energy of 10$^{51}$ ergs, the percentage of events for which CTA allows us to cover the 90 \% C.R. of the GW skymap is $\sim$ 60 \% and 73 \% for a cut-off energy of 30 GeV and 100 GeV respectively. For the lowest energetic events, this percentage is   $<$1 \% (cut-off at 30 GeV) and 1.5 \% (cut-off at 100 GeV), see Tab. \ref{tab:coverage} for details. 

In Fig. \ref{fig:cumarea} it is also shown the CTA coverage of GW skymaps as a function of the distance of the BNS systems. It can be seen that, for GRBs as energetic as GRB 090510, for almost all the sources the coverage is 90 \% C.R. independently on the distance; instead, for the less energetic GRBs, the coverage decreases by increasing the distance of the sources, as expected. For instance, for GRBs with E$_{\rm iso}=10^{51}$ ergs and cut-off energy of 100 GeV, the coverage starts decreasing from the 90 \% C.R. when the sources are located at distances $>$ 200 Mpc. This information can be used to further optimize the observational strategy when information about the distance of the source is available in the GW alerts. 

\begin{figure}[h!]
\begin{center}
\includegraphics[scale=1]{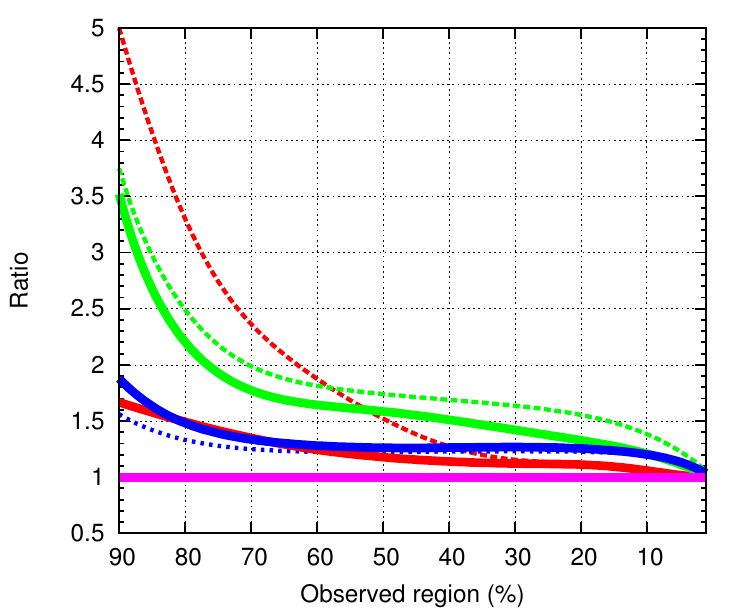}
\caption{Ratio between the CTA coverage with consecutive observations estimated with the strategy proposed in this work and using a constant observing time for each field. The different colours refer to different assumed values for the isotropic energy of the simulated short GRB: 3.5 10$^{52}$ erg (magenta), 10$^{51}$ erg (blue), 10$^{50}$ erg (green) and 10$^{49}$ erg (red); the dotted (solid) lines refer to a cut-off energy of 100 GeV (30 GeV).}\label{fig:ratio}
\end{center}
\end{figure}

We also estimated, for each simulated event, the percentage of the GW skymap that can be covered with consecutive CTA observations, by assuming a different strategy: we assumed a constant observing time for each observed field as done, for instance, in \cite{2014MNRAS.443..738B}. To make a direct comparison with the observational strategy proposed in this work, for each event we consider the same total observing time. The results are shown in Fig. \ref{fig:ratio}. It can be seen that, while for the most energetic events the results obtained with the two strategies are compatible, for lower energies with the strategy proposed in this work there is a significant improvement in the coverage (and therefore in the detection rates). For instance, for sources with E$_{\rm iso}$=10$^{50}$ ergs and a cut-off energy of 100 GeV, the percentage of events for which CTA allows us to cover the 90 \% C.R. of the GW skymap increases by a factor of $\sim$ 3.7. This translates into an increase in the joint GW and EM detection rate by a factor of $\sim$ 2. We note that the improved results is a consequence of the knowledge on the source parameters.

\begin{table}[tbp]    
\begin{center}
\scalebox{0.88}{
\begin{tabular}{lllcccc}
\hline \hline   \vspace{0.1mm}\\
E$_{\rm iso}$ & & cut-off & \% of events & \% of events\vspace{1mm}\\ 
(ergs) & & (GeV) & Obs. region = 90\% & Obs. region  $\geq$ 50 \% \vspace{1mm}\\ 
\hline
\vspace{1mm}\\  
 10$^{49}$& & 30 & $<$ 1& $<$ 1 \vspace{1mm}\\ 
  &  &  100 &  1.5 & 1.9  \vspace{1mm}\\ 
  \hline \vspace{1mm}\\  
   10$^{50}$ &  & 30 & 8.8 & 12.2 \vspace{1mm}\\ 
 &   & 100 &  18.0 & 28.8 \vspace{1mm}\\ 
  \hline \vspace{1mm}\\  
 10$^{51}$ &  & 30 & 59.7 & 74.5 \vspace{1mm}\\ 
 &   & 100 & 73.0 & 85.1  \vspace{1mm}\\ 
  \hline \vspace{1mm}\\  
3.5$\times  10^{52}$ &  & 30 & 99.9 & 100 \vspace{1mm}\\ 
 &   & 100 & 99.9 & 100  \vspace{1mm}\\ 
\hline\hline
\end{tabular}
}
\caption{Percentage of GW events for which CTA will allow to cover an area of the GW skymaps equal to 90 \% C.R. and $> $ 50 \% C.R.} \label{tab:coverage}

\end{center}
\end{table}


\subsection{Joint GW and EM detection rates}\label{sec:rates}
We estimated the joint GW and EM detection rates. To do this, for each event and for each set of parameters describing the EM emission, we checked the percentage of the GW error box that can be covered by CTA (see Secs. \ref{sec:strategy} and \ref{ctadetection}); from this we got a list of CTA pointings and a list of corresponding observing times. Then, we verify if the simulated source is inside at least one of the field that can be observed with CTA. The sensitivity adopted to estimate the percentage of GW error box that can be covered refer to on-axis sources; however, according to \cite{2013APh....43..317D} the sensitivity remains uniform within a radius of $\sim$ 1 deg. Furthermore, we verified that, for a radius up to 1 deg, the distribution of statistical significance is consistent with the one obtained for an on-axis source, so the estimated observing time is still sufficient to guarantee the EM detection of the source. Therefore, we considered a source as detected by CTA if it is located within a radius from the pointing coordinates of 1 deg. 

Once we got the number of GW detected events that can potentially be observed by CTA, we estimated the joint GW and EM detection rates. For simplicity, we assumed that each simulated source is visible for CTA for the overall observing time estimated with our strategy, independently on its location and on the time of the GW trigger; we also fixed the zenith angle of the GRBs to 20$^\circ$ (we recall that the public IRFs used in this work have been generated assuming $\theta_z$=20$^\circ$, see Sec. \ref{sec:obstime}). Then, the joint detection rates have been estimated taking into account the CTA duty cycle, that we conservatively assumed to be of $\sim$ 10\% \footnote{For instance, the maximal duty cycle of the MAGIC telescope considering only moonless night is $\sim$ 18\% (see e.g. \cite{2017APh....94...29A}); the actual duty cycle is however lower that this value, due to several factors that should be taken into account such as, for example, the weather conditions, technical operations on the instruments etc.}. The results are shown in Table \ref{tab:jointGWEM}.


\begin{table}[tbp]    
\begin{center}
\scalebox{0.88}{
\begin{tabular}{lllccc}
\hline \hline   \vspace{0.1mm}\\
E$_{\rm iso}$ & cut-off & EM and GW \vspace{1mm}\\  
 (ergs) & (GeV) & (yr$^{-1}$) \vspace{1mm}\\ 
\hline
\vspace{1mm}\\  
 10$^{49}$& 30 & $<$ 10$^{-3}$  \vspace{1mm}\\ 
  & 100 & 0.001 \vspace{1mm}\\ 
  \hline \vspace{1mm}\\  
   10$^{50}$ & 30 & 0.01 \vspace{1mm}\\ 
  & 100 &  0.03 \vspace{1mm}\\ 
  \hline \vspace{1mm}\\  
 10$^{51}$ & 30 & 0.06 \vspace{1mm}\\ 
  & 100 & 0.07 \vspace{1mm}\\ 
  \hline \vspace{1mm}\\  
3.5$\times  10^{52}$ & 30 & 0.08 \vspace{1mm}\\ 
  & 100 & 0.08 \vspace{1mm}\\ 
\hline\hline
\end{tabular}
}
\caption{Expected rates of joint EM and GW detections.}\label{tab:jointGWEM}

\end{center}
\end{table}

It can be seen that, when considering the most energetic short GRBs, the joint GW and VHE EM detection rates is 0.08 yr$^{-1}$. This value should be considered as the upper limit to the true detection rate, since short GRBs are expected to have a distribution of possible isotropic energies within the range considered in this work; furthermore, this estimate has been obtained under the optimistic assumption that all BNS mergers are progenitors of short GRBs with VHE afterglow emission. As mentioned in the Introduction, for GRB 170817A no VHE emission has been detected; by extrapolating the light curve and spectrum observed by \emph{Fermi}-GBM \cite{2017ApJ...848L..14G} to VHE, the associated VHE flux would be too low in order to be detected with CTA. However, the presence of an extra, additional VHE component cannot be excluded, since VHE observations started with a significant time delay with respect to the GRB onset (see Sec. \ref{sec:intro}).


The results we found strictly depends also on the assumed BNS merger rate and on the assumed jet opening angle of short GRBs. By re-scaling the values for the current estimates of the BNS merger rates \cite{2017PhRvL.119p1101A}, the estimated joint detection rate for the most energetic short GRBs is in the range (0.03 - 0.5) yr$^{-1}$. The rates may further increase if observations are performed also during moonlight: in this case the duty cycle is expected to increase up to a factor $\sim$ 2, although a sensitivity degradation is also expected and may affect the number of detections (see, e.g., \cite{2012APh....35..435A,2017APh....94...29A}). The joint GW and EM detection rate may be higher also if sub-threshold GW events with lower signal-to-noise ratio with respect to the one used in this work \cite{2016JCAP...11..056P} are EM followed-up. However, for this kind of events the GW sky localization is expected to be larger, making it more difficult the EM follow-up; furthermore, the possibility that the sub-threshold GW signal is a false positive GW candidate should be taken into account. Finally, the joint GW and EM detection rates are expected to increase when KAGRA \cite{2012CQGra..29l4007S,2013PhRvD..88d3007A} and LIGO-India \cite{LigoIndia} will join the GW detector network (see \cite{2017arXiv1304.0670L}) and if also off-axis short GRBs and/or larger values for $\theta_j$ are considered (see Sec. \ref{sec:uncertainties}); a detailed study of the off-axis emission in the VHE energy range is beyond the scope of the paper.

\subsection{Uncertainties on the joint EM and GW detection rates}\label{sec:uncertainties}
The detection rates estimated in Sec. \ref{sec:rates} depends on the assumptions on the parameters describing the EM emission, for instance on the spectral index and on the temporal decay index of the flux. Furthermore, while we assumed a ``fiducial'' $\theta_j$=10$^\circ$, other values of the jet opening angle are also possible. We estimated the variations in the detection rates associated to these uncertainties.

The spectral index that we used in this work, $\beta=$-2.1, has an associated uncertainty of 0.1 \cite{2010ApJ...709L.146D}. We estimated the associated variation in the detection rates by varying the value of $\beta$ of the simulated GRBs from -2.0 to -2.2. We found that, for the less energetic GRBs (E$_{\rm iso}=10^{49}$ ergs), the joint GW and EM detection rate can increase by a factor of 3 for $\beta$=-2.0, while it stays < $10^{-3}$ for $\beta$=-2.2; for the most energetic GRBs (E$_{\rm iso}=3.5\times 10^{52}$ ergs) the joint detection rate changes by less than 1\%. The overall range of detection rate is the same estimated for $\beta=2.0$: $< 10^{-3}$ yr$^{-1}$ - 0.08 yr$^{-1}$.

Also the temporal decay index of the GRB light curve has an associated uncertainty; specifically, in this work we used the value reported in \cite{2016JCAP...11..056P}: $\delta$=1.60 $\pm$ 0.15. We estimated the variations in the detection rate associated to the uncertainty in the temporal decay index of the GRB light curve by varying the value of $\delta$ from 1.45 to 1.75 (see also Sec. \ref{sec:emmodel}). Also in this case, we found the detection rates changes only for the less energetic GRBs. For instance we found that, for events with E$_{\rm iso}=10^{49}$ ergs, the detection rates increase by a factor $\sim$ 8, while it stays < $10^{-3}$ for $\delta$=1.75. The overall range of detection rate is the same as estimated for $\delta$=1.60.

Finally, we estimated how the detection rates change depending on the jet opening angle $\theta_j$. As already said in Sec. \ref{sec:emmodel}, in this work we used a fiducial $\theta_j=10^{\circ}$, however, there are large uncertainties on its value. The value of $\theta_j$ is usually inferred from observation of a break in the afterglow light curve\footnote{When $\Gamma^{-1}=\theta_j$ a steepening in the flux decay of the afterglow emission is expected to be observed (the so called ``jet break'', see e.g. \cite{2004RvMP...76.1143P}).}; also, the lack of such a break is used to put lower limits on $\theta_j$. The lowest opening angle is the one inferred for GRB 090510, estimated to be between $\sim$ 0.1$^{\circ}$ and 1$^{\circ}$ \citep{2010ApJ...720.1008C,2010MNRAS.409..226K,2011MNRAS.414.1379P}. The highest lower limit for the opening angle is the one estimated for GRB 050724: this burst has no observed break after 22 days, leading to $\theta_j >  25^{\circ}$ \citep{2012MNRAS.425.2668C}. Finally, there are numerical studies suggesting that $\theta_j \leq 30^\circ$ (see, e.g., \cite{2011ApJ...732L...6R}). We therefore estimated the detection rates considering  $1^\circ \leq \theta_j \leq 30^{\circ}$. We found that, for $\theta_j=1^\circ$, the detection rates decrease by a factor 0.1, while for $\theta_j=30^\circ$ the detection rates increase by a factor $\sim$ 9. Therefore, the overall joint GW and VHE EM detection rate is in the range  ($< 10^{-3}$ - 0.7) yr$^{-1}$.


\section{Discussion and conclusions}\label{sec:concl}
We presented a detailed study on the expectations for joint GW and VHE EM observations of BNS mergers with the interferometers Advanced Virgo and Advanced LIGO and with the VHE gamma-ray observatory CTA. We used detailed EM and GW simulations of the emitted GW and EM signals and their detection, and along with the assumed source parameters we proposed an optimized observational strategy for CTA EM follow-up of GW events, to increase the chance of detecting the VHE EM counterparts. We showed that, when considering the most energetic short GRBs, CTA will be able to cover the 90 \% C.R. of the GW skymap  for $\sim 100$ \% of the GW events; this percentage is lower (between $\sim$ 60 \% and $\sim$ 70\%, depending on the assumed cut-off energy) when GRBs with intermediate energies are considered. The associated expected rates of joint GW and VHE EM detections are 0.08 yr$^{-1}$ if all the simulated short GRBs are as energetic as GRB 090510, $< 10^{-3}$ yr$^{-1}$ if only lowest energetic GRBs are considered. 
The detection rate of GRBs with E$_{\rm iso}$ < $10^{49}$ erg would be negligible.
However, it is important to point out that these estimates are conservative: much higher detection rates are expected if higher BNS merger rate are assumed, as described in Sec.\ref{sec:results}. A further increase is expected if CTA will be able to observe also during low-moonlight conditions. 

We have also shown that the proposed strategy is more efficient to detect intermediate energetic GRBs with respect to more common observational strategies assuming the same observing time for each observed field, provided that the source parameters are known. For instance, for sources with E$_{\rm iso}$=10$^{50}$ ergs and cut-off energy at 100 GeV, the joint GW and EM detection rate increases by a factor of $\sim$ 2.

The results here presented have shown that the EM follow-up observations at VHE with CTA represent a promising instrument to identify the VHE EM counterpart of GW transient events detected by Advanced Virgo and Advanced LIGO. It is also important to point out that the observational strategy here proposed depends on the prior assumption on the source properties (light curve and spectrum), but can be generalized to other EM emission models and other telescopes. 
Different options on the choices of the source parameters can be investigated and implemented according to prior knowledge obtained by the GW event or from assumed properties of the GRB.
 In case of a joint GW and EM trigger, for instance by \emph{Fermi}-GBM or INTEGRAL (as occurred in the case of GW170817/GRB170817A), the EM alert sent to the astronomical community could also contain information about E$_{\rm iso}$, that can be used to model the VHE emission. If this information is not available, different observational schemes can be envisaged in our approach and can be treated as sub-cases of the general framework described in the paper. For instance, one of the outcomes of our work is the GW sky localization areas that can be covered with CTA consecutive observations, as a function of the distance of the BNS systems (Fig.\ref{fig:cumarea}, right panel). If we know the distance of the source (for instance from a GW alert), we can derive the minimum brightness that a GRB should have to be detected with CTA and thus use this brightness to compute the observing times for each CTA pointing. 
 Another option is, for instance, to assume that only the most energetic GRBs (such as GRB 090510) also have an extended VHE emission, and optimize the observational strategy to search for this kind of sources. Finally, the strategy here proposed can be easily adapted to take into account the distribution of galaxies in the local universe: this will help to reduce the tiling effort considerably, increasing the chance of detecting the EM counterpart at least of the closest GW events. 

This work represents an important step describing the potential of joint GW and VHE EM observations. A comparison of the future observations with the joint GW and VHE EM detection rates here estimated could help to shed light on the physics of compact objects and on the acceleration processes. 

\appendix
\section{TS map and post trial significance}\label{sec:TS}
To evaluate if a source is detected or not, the likelihood ratio Test Statistic (TS) can be used. The TS is twice the logarithm of the ratio of the likelihood L$_1$ evaluated at the best-fit model parameters when including a candidate point source at a given position to the likelihood L$_0$ evaluated at the best-fit parameters under the baseline (no source) model:
\begin{equation}
\rm{TS}=2 \left( \log L_1 - \log L_0\right).
\end{equation} 

Since, when receiving a GW alert, we don't know with high level of accuracy the actual position of the source, to detect the source one can be to produce a TS skymap: the specified source is displaced on a grid of sky directions and, for each direction, the TS value is computed. Specifically, in this work we considered a total monitored area equal to the CTA-LST FoV (4.5$^\circ$), we divided it into several pixels and then we estimate the TS values assuming that the source is located at the center of the pixels. Each pixel corresponds to a different trial; we call n$_{\rm trials}$ the total number of pixels of the map (i.e., the total number of trials).

The TS for a single pixel is distributed as $\frac{1}{2} \chi^2(x)$ with 1 degree of freedom \cite{1996ApJ...461..396M}; the factor $\frac{1}{2}$ comes from the fact that we force the point source to have a positive flux. Therefore, the probability of obtaining a value of TS which exceeds a given threshold TS$_0$ in a single pixel is
\begin{equation}
P_1=\frac{1}{2} \int_{\rm TS_0}^\infty \chi_1(x)^2 dx.
\end{equation}
As a basic rule of thumb, the significance of an excess can be estimated as $\sigma=\sqrt{{\rm TS}}$.

We consider a pixel angular size of 0.2$^\circ$, that roughly corresponds to the PSF of CTA at $\sim$50 - 100 GeV: in this way we can reasonably assume that the signals measured in the various pixels are not correlated (i.e., they are statistically independent). Therefore, the probability P$^{'}$ of obtaining TS > TS$_0$ 1 time after testing all pixels can be described as a  binomial distribution: 

\begin{equation} \label{eq:prob}
P^{'}=\binom{\rm{n_{\rm trials}}}{1} P_1 (1-P_1)^{\rm{n_{trials}-1}}\sim1-(1-P_1)^{n_{\rm trials}} \sim P_1*n_{\rm trials}
\end{equation}

We also have (see, e.g., \cite{2013A&A...552A.134D})
\begin{equation}\label{eq:prob2}
P^{'}=\frac{1}{2} \int_{\rm TS_0^{'}}^\infty \chi_1(x)^2 dx
\end{equation}

For a post-trial significance threshold S$_0^{'}$=$\sqrt{\rm TS_0^{'}}$ = 5 $\sigma$, the corresponding value of $TS_0$ (and so of S$_0$) can be obtained by equating eqs. \ref{eq:prob} and \ref{eq:prob2}, for a given n$_{\rm trials}$. 
In this work we considered n$_{\rm trials} \sim$ 500, that corresponds to the minimum number of pixels of a TS map enclosing the LST FOV (we don't consider the larger MST FOV, since the CTA sensitivity is expected to decrease for angular distances greater than 1 deg from the pointing coordinates). Therefore, in this work we use S$_0 \sim$ 6.


\acknowledgments
We thank F. Schussler for useful discussions and comments. The work was made possible thanks to the efforts of the CTA consortium in providing detailed instrument response function for the CTA array.
We acknowledge support from the Scuola Normale Superiore, Pisa (Italy) through the funding of the project "Electromagnetic Follow-up of Gravitational Wave Sources: Identification and Characterization of their Counterparts via Multi-Wavelength Observations", cod.SNS16\_B\_STAMERRA.
This research made use of \verb|ctools|, a community-developed analysis package for IACT data. \verb|ctools| is based on GammaLib, a community-developed toolbox for the high-level analysis of astronomical gamma-ray data.
This paper has been approved for publication by the LIGO Scientific Collaboration (LIGO Document P1700441) and the Virgo Collaboration (Virgo Document VIR-0917A-17). This paper has gone through internal review by the CTA Consortium.

\bibliographystyle{JHEP}
\bibliography{EMFup,GW-CTA-paper_bibliography}




\end{document}